\documentclass[prl,reprint,amsmath,amssymb]{revtex4-1}
\usepackage{graphicx}
\usepackage{dcolumn}
\usepackage{bm}
\usepackage{epsfig}
\usepackage{amssymb}
\usepackage{amsmath}

\usepackage{color}

\begin{document}

\title{Bond percolation thresholds on Archimedean lattices from critical polynomial roots}
\author{Christian R. Scullard$^1$}
\affiliation{$^1$Lawrence Livermore National Laboratory, Livermore, California, USA}
\author{Jesper Lykke Jacobsen$^{2,3,4}$}
\affiliation{$^2$Laboratoire de Physique de l'Ecole Normale Sup\'erieure, ENS, Universit\'e PSL, CNRS, Sorbonne Universit\'e, Universit\'e de Paris, Paris, France}
\affiliation{$^3$Sorbonne Universit\'e, \'Ecole Normale Sup\'erieure, CNRS, Laboratoire de Physique (LPENS), 75005 Paris, France} 
\affiliation{$^4$Institut de Physique Th\'eorique, Universit\'e Paris Saclay, CEA, CNRS, 91191 Gif-sur-Yvette, France}

\date{\today}

\begin{abstract}
We present percolation thresholds calculated numerically with the eigenvalue formulation of the method of critical polynomials; developed in the last few years, it has already proven to be orders of magnitude more accurate than traditional techniques. Here we report the result of large parallel calculations
to produce what we believe may become the reference values of bond percolation thresholds on the Archimedean lattices for years to come. For example, for the kagome lattice we find $p_{\rm c}=0.524\,404\,999\,167\,448\,20 (1)$, whereas the best estimate using standard techniques is $p_{\rm c}=0.524\,404\,99(2)$.
We further provide strong evidence that there are two classes of lattices: one for which the first three scaling exponents characterizing the finite-size corrections
to $p_{\rm c}$ are $\Delta=6,7,8$, and another for which $\Delta=4,6,8$.
We discuss the open questions related to the method, such as the full scaling law, as well as its potential for determining critical points of other models.
\end{abstract}

\maketitle

\begin{figure}
\begin{center}
 \includegraphics[width=6.5cm]{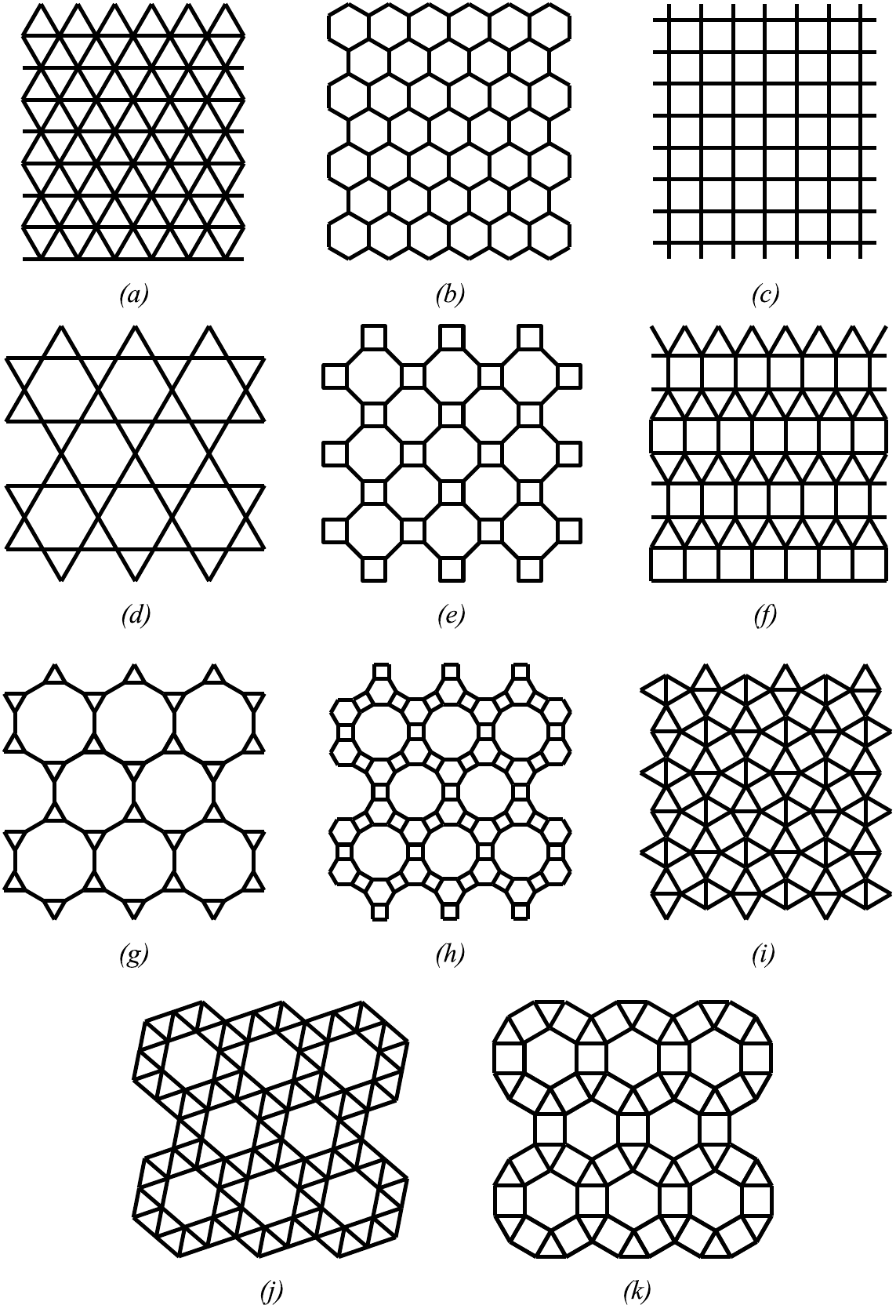}
 \caption{The eleven Archimedean lattices.}
 \label{fig:archi}
\end{center}
\end{figure}

\paragraph{Introduction.}
Percolation is one of the simplest mathematical models with a phase transition \cite{BroadbentHammersley57}. It has served as a paradigm of such models, with basic properties that also emerge in a diverse range of systems from superconductivity \cite{Bardeen} to black hole critical collapse \cite{Choptuik}. In recent years the two-dimensional problem, which we focus on here, has proven to be particularly interesting, with its fascinating mix of solved and unsolved problems. Given a lattice, such as one of those shown in Fig. \ref{fig:archi}, choose each bond to be open with probability $p$ and closed with probability $1-p$ independently of all others. As $p$ is increased, there is a critical threshold, $p_{\rm c}$, which marks a sharp transition to a regime with an infinite connected cluster. Despite the problem's apparent simplicity, the value of $p_{\rm c}$ is unknown for most lattices, with exact solutions available only on a restricted class \cite{SykesEssam, Wierman84, Scullard06, Ziff06, ZiffScullard06, BollobasRiordan11}. There have been great advances in the understanding of the continuum limit using conformal invariance and stochastic Loewner evolution \cite{Cardy92,SaleurDuplantier87,Schramm2001,Smirnov}, in which the details of the underlying lattice are irrelevant, but progress on unsolved lattice-dependent quantities, such as the critical probabilities of most of the Archimedean lattices, has been limited to the derivation of rigorous bounds \cite{MayWierman05,Wierman03,Wierman2002b,Wierman15,MayWierman07,RiordanWalters07,Wierman2017} (though these are ever-tightening) and numerical studies \cite{FengDengBlote08,Parviainen,SudingZiff99}. For example, the critical probability for bond percolation on the kagome lattice is known rigorously to satisfy \cite{Wierman2017}
\begin{equation}
 0.522 551 < p_c < 0.526 490,
\end{equation}
a range with a width of $10^{-3}$, and the best estimate using traditional numerical techniques is $p_c = 0.52440499(2)$ \cite{FengDengBlote08}, an accuracy of $10^{-8}$. It is very possible that in order to gain a complete understanding of subjects like conformal invariance and universality, a solution to, or at least a firmer grasp of, these lattice-specific problems will be necessary. In this Letter, we make progress towards this objective by pushing the precision of $p_{\rm c}$ to the order of $10^{-18}$ in the most favorable case.

\paragraph{Critical polynomials.}
The method of critical polynomials originated from the observation that in all exactly-solved problems $p_{\rm c}$ is the (unique) root, $p_{\rm c} \in (0,1)$, of a polynomial. For example on the triangular lattice (Fig. \ref{fig:archi}a), the bond percolation threshold is given by \cite{SykesEssam}
\begin{equation}
 p_{\rm c}^3-3p_{\rm c}+1=0 \,, \label{eq:tricrit}
\end{equation}
so that $p_{\rm c}=2 \sin \pi/18 \approx 0.347296$. Similar results are obtained for the hexagonal (Fig.\ \ref{fig:archi}b) and square (Fig.\ \ref{fig:archi}c) lattices. This polynomial can be generalized unambiguously even to problems for which the exact solution is not known \cite{ScullardZiff08, ScullardZiff10, Scullard11-2, ScullardJacobsen2012}. This is done by first choosing a finite subgraph, $B$, called the basis, that generates the infinite lattice when copies are arranged in some periodic way. Next, assuming the percolation realization is identical on each copy of the basis, we use the label $\mathrm{2D}$ for the event that there is an open cluster connecting every copy of $B$ and $\mathrm{0D}$ for the event that no infinite set of bases can be connected by open clusters. Denoting the probabilities of these events $Z_p(\mathrm{2D})$ and $Z_p(\mathrm{0D})$, the critical polynomial $P_B(p)$ is defined by
\begin{equation}
 P_B(p) \equiv Z_p(\mathrm{2D})-Z_p(\mathrm{0D}) \,. \label{eq:P_B}
\end{equation}
This is clearly a polynomial in $p$ as $B$ has a finite number of edges. For reasons related to universality \cite{ScullardJacobsen2012}, the root of the polynomial in $[0,1]$ provides an estimate of the critical point that becomes more accurate as the size of $B$ is increased. For problems with an exact solution, remarkably, the critical polynomial provides the correct answer for any choice of $B$---even the smallest possible---and this convenient property has been used to find some previously unknown exact solutions \cite{ScullardJacobsen2012,ScullardZiff10}.

\paragraph{Transfer matrix.}
The origins, development and refinement of the method can be followed in a series of papers written over the last several years  \cite{ScullardZiff08,ScullardZiff10,Scullard11-2,Jacobsen12,ScullardJacobsen2012,Jacobsen14b,Jacobsen15,ScullardJacobsen16}. Although different methods have been used to compute $P_B$ \cite{ScullardZiff10,Scullard11-2,Scullard11}, the transfer matrix \cite{ScullardJacobsen2012,Jacobsen14} has proven to be by far the most efficient. It is in fact not really necessary to compute the polynomial explicitly if all one wants is the root, and in Ref.\ \cite{Jacobsen15} it was shown by one of us that at this root, the largest eigenvalues of two transfer matrices, $T_{\mathrm{open}}(p)$ and $T_{\mathrm{closed}}(p)$, are equal. These two transfer matrices describe different topological sectors and build up respectively $Z_p(\mathrm{2D})$ and $Z_p(\mathrm{0D})$ on a semi-infinite cylinder of width $n$ lattice spacings (the reader is referred to Ref.\ \cite{Jacobsen15} for more details). The strategy is then to start with some value of $p$ close to $p_c$ and continually compute the largest eigenvalues $\Lambda_{\mathrm{open}}$ and $\Lambda_{\mathrm{closed}}$, adjusting $p$ until
\begin{equation}
 \Lambda_{\mathrm{open}}=\Lambda_{\mathrm{closed}}
\end{equation}
to within some tolerance. The great advantage of this method is that it allows bases of size as large as $n=12$ to be tractable on an ordinary desktop, with extrapolation used to find the estimate on the infinite lattice.

Here, we use a parallel implementation of the method, which is run on supercomputers, to reach $n=16$ for bond percolation on the unsolved Archimedean lattices. These are the eleven two-dimensional lattices for which all vertices are equivalent (see Fig.\ \ref{fig:archi}). The square, triangular and hexagonal lattices have exact solutions, such as eq.~(\ref{eq:tricrit}), but the remaining eight are unsolved. Our results for those are shown in Table \ref{tab:pc}, along with the best estimates using Monte Carlo \cite{Parviainen} or traditional transfer matrix techniques \cite{FengDengBlote08}. 

\begin{table}
\begin{center}
 \begin{tabular}{lr|l|lr}
 Lattice & & $p_{\rm c}$ (this work) & $p_{\rm c}$ (numerical) & Ref. \\ \hline
 Kagome        & $(d)$ & 0.524\,404\,999\,167\,448\,20(1) & 0.524\,404\,99(2) & \cite{FengDengBlote08} \\
 Four-eight      & $(e)$ & 0.676\,803\,124\,390\,011\,3(3) & 0.676\,802\,32(63) & \cite{Parviainen} \\
 Frieze            & $(f)$ & 0.419\,640\,358\,863\,69(2) &  0.419\,641\,91(43) & \cite{Parviainen} \\
 Three-twelve   & $(g)$    & 0.740\,420\,798\,850\,811\,610(2) &  0.740\,421\,95(80) & \cite{Parviainen} \\
 Cross             & $(h)$ & 0.693\,733\,124\,922(2) &  0.693\,733\,83(72) & \cite{Parviainen} \\ 
 Snub square    & $(i)$    & 0.414\,137\,856\,591\,7(1) &   0.414\,137\,43(46) & \cite{Parviainen} \\
 Snub hex.  & $(j)$   & 0.434\,328\,317\,224\,0(6) &  0.434\,328\,0(5) & \cite{ZiffComm} \\ 
 Ruby            & $(k)$   & 0.524\,831\,461\,573(1) &  0.524\,832\,58(53) & \cite{Parviainen} \\
 \end{tabular}
 \caption{Percolation thresholds for the nine unsolved Archimedean lattices, as compared to their previous numerical determinations. The letters refer to Fig.\ \ref{fig:archi}.}
 \label{tab:pc}
\end{center}
\end{table}

\paragraph{Iterative scheme.}
To find the eigenvalues for a given value of $p$, we use the power iteration method, which involves simply multiplying an initial vector repeatedly by the transfer matrix, until it converges to the eigenvector with the largest eigenvalue. We demand convergence to 40 decimal places, which requires the use of an arbitrary precision library \cite{cln}. Note that the computations are actually done using $v \equiv p/(1-p)$, the bond weight in the Fortuin-Kasteleyn representation of the corresponding $Q=1$ state Potts model \cite{FK1972}. Once we have the values of $\Lambda_{\mathrm{open}}(v)$ and $\Lambda_{\mathrm{closed}}(v)$, we use them to find the next value of $v$, and so on until the two eigenvalues are identical within our tolerance of $\delta \equiv 10^{-40}$. To adjust $v$, we use Householder's method of order 2 \cite{Householder}. That is, if the equation to solve is $f(v)=0$, then the $m^{\mathrm{th}}$ iteration is given by
\begin{equation}
v_{m+1}=v_m+2 \frac{\left(\frac{1}{f}\right)'(v_m)}{\left(\frac{1}{f}\right)''(v_m)} \,. \label{eq:Householder}
\end{equation}
For our purposes, $f(v)=\Lambda_{\mathrm{open}}(v)-\Lambda_{\mathrm{closed}}(v)$. To approximate the derivatives we use the central difference formulas
\begin{eqnarray}
 f'(v) &\approx& \frac{g_2-g_0}{2 \epsilon} \, \\
 f''(v) &\approx& \left(\frac{g_2-g_1}{\epsilon}-\frac{g_1-g_0}{\epsilon} \right)\epsilon^{-1} \,,
\end{eqnarray}
where we set $\epsilon = \sqrt{\delta}$, and
\begin{eqnarray}
 g_0 &\equiv& f(v-\epsilon) \,, \\
 g_1 &\equiv& f(v) \,, \\
 g_2 &\equiv& f(v+\epsilon) \,.
\end{eqnarray}


One typically needs three or four Householder iterations to get $v$ converged to the target precision $\delta$. We take care to choose the initial $v$
carefully, using extrapolations of the $v_{\rm c}$ obtained for smaller $n$. Indeed, if $|v-v_{\rm c}| \ll 1$ each second-order Householder iteration
doubles the number of correct digits. Arriving at our largest size, $n=16$, we are most often able to pick the initial $v$ correct to about $10^{-18}$,
so a single iteration gives sufficient precision for our purposes. Only in the case of the snub hexagonal lattice (Fig.\ \ref{fig:archi}j), we performed two iterations for $n=16$ because we found our initial guess to be insufficiently accurate.

\paragraph{Parallelization.}
The parallel algorithm is used to compute $\Lambda_{\mathrm{open}}$ and $\Lambda_{\mathrm{closed}}$ at $v$ and $v \pm \epsilon$, by the actual transfer matrix multiplication. It will be described more fully elsewhere, but in broad terms the goal is to distribute the components of the vector across processors in such a way that the need for inter-processor communication is minimized. Our implementation was inspired by Jensen's \cite{Jensen03} transfer matrix enumeration of self-avoiding polygons. In that problem, Jensen identified a criterion for organizing the vector components that ensured that any transfer operation performed on data on a particular processor would not require information from any other. In our problem, as far as we know, it is not possible to replicate this exactly, but using a variant of Jensen's approach we were able to keep the need for communication to a minimum.

Up to $n=12$, calculations can be done on ordinary desktops. Even then, the accuracy achieved by the method is far better than that of traditional techniques \cite{Jacobsen15}; our calculations up to $n=16$, should place these quantities permanently out of their reach. For the $n=16$ computations, we used
$\simeq 10^3$ processors.
The time needed to complete a single power iteration varies with the lattice, but typically takes three to four hours. The number of power iterations needed to get convergence of a single eigenvalue is likewise variable and depends strongly on the initial vector used. When doing the first Householder iteration, we start with a vector with only one non-zero component. In this case, the number of power iterations needed is in the 40--60 range. However, for subsequent Householder iterations, we start with the final vector computed during the previous iteration and in this way we can reduce the number of power iterations needed to the 20--30 range as $p$ approaches $p^*$. 

\begin{table}
\begin{center}
 \begin{tabular}{l|ll}
 $n$ & $p_{\rm c}$ \\ \hline
 1 & 0.524\,429\,717\,521\,274\,793\,546\,879\,681\,534\,455\,071\,6205 \\
 2 & 0.524\,406\,057\,896\,062\,634\,245\,378\,836\,666\,345\,666\,7920 \\
 3 & 0.524\,405\,092\,218\,718\,391\,406\,491\,710\,278\,995\,604\,5159 \\
 4 & 0.524\,405\,013\,882\,343\,450\,677\,924\,933\,274\,891\,201\,3263 \\
 5 & 0.524\,405\,002\,666\,098\,533\,997\,468\,638\,043\,799\,737\,1046 \\ 
 6 & 0.524\,405\,000\,252\,138\,641\,166\,065\,238\,385\,312\,009\,4089 \\
 7 & 0.524\,404\,999\,570\,802\,604\,857\,648\,689\,641\,603\,340\,3853 \\
 8 & 0.524\,404\,999\,338\,748\,706\,184\,041\,906\,677\,709\,350\,6317 \\
 9 & 0.524\,404\,999\,247\,980\,209\,806\,701\,838\,958\,679\,653\,4330 \\
 10 & 0.524\,404\,999\,208\,475\,451\,262\,855\,277\,119\,432\,089\,6813 \\
 11 & 0.524\,404\,999\,189\,755\,973\,511\,801\,309\,010\,812\,928\,2307 \\
 12 & 0.524\,404\,999\,180\,248\,443\,779\,969\,638\,346\,824\,671\,7858 \\
 13 & 0.524\,404\,999\,175\,132\,845\,053\,820\,303\,018\,487\,609\,0453 \\
 14 & 0.524\,404\,999\,172\,242\,908\,087\,780\,703\,763\,071\,248\,1530 \\
 15 & 0.524\,404\,999\,170\,540\,780\,670\,080\,646\,173\,449\,291\,2196 \\
 16 & 0.524\,404\,999\,169\,501\,410\,335\,190\,170\,832\,654\,998\,6109 \\
 \hline
 $\infty$ & 0.524\,404\,999\,167\,448\,20 (1) \\
 \end{tabular}
 \caption{Bond percolation thresholds for the kagome lattice computed on semi-infinite cylinders of width $n$.}
 \label{tab:kagome}
\end{center}
\end{table}

\paragraph{Extrapolation.}
Our extrapolation scheme for the resulting sequences is based on the empirical scaling form
\begin{equation}
  p_{\rm c}(n)=p_{\rm c}+\sum_{k=1}^\infty \frac{A_k}{n^{\Delta_k}} \,, \label{scaling-form}
\end{equation}
where the amplitudes $A_k$ and exponents $\Delta_k$ are dependent upon the lattice. There is currently little theoretical understanding of this scaling, and we compute the $A_k$ and $\Delta_k$ by simply fitting the actual data. Fortunately, we do now have a fair number of data points;
in Table \ref{tab:kagome}, we list the thresholds computed for $n \le 16$  for the kagome lattice.

\begin{figure}
\begin{center}
 \includegraphics[width=4.0cm]{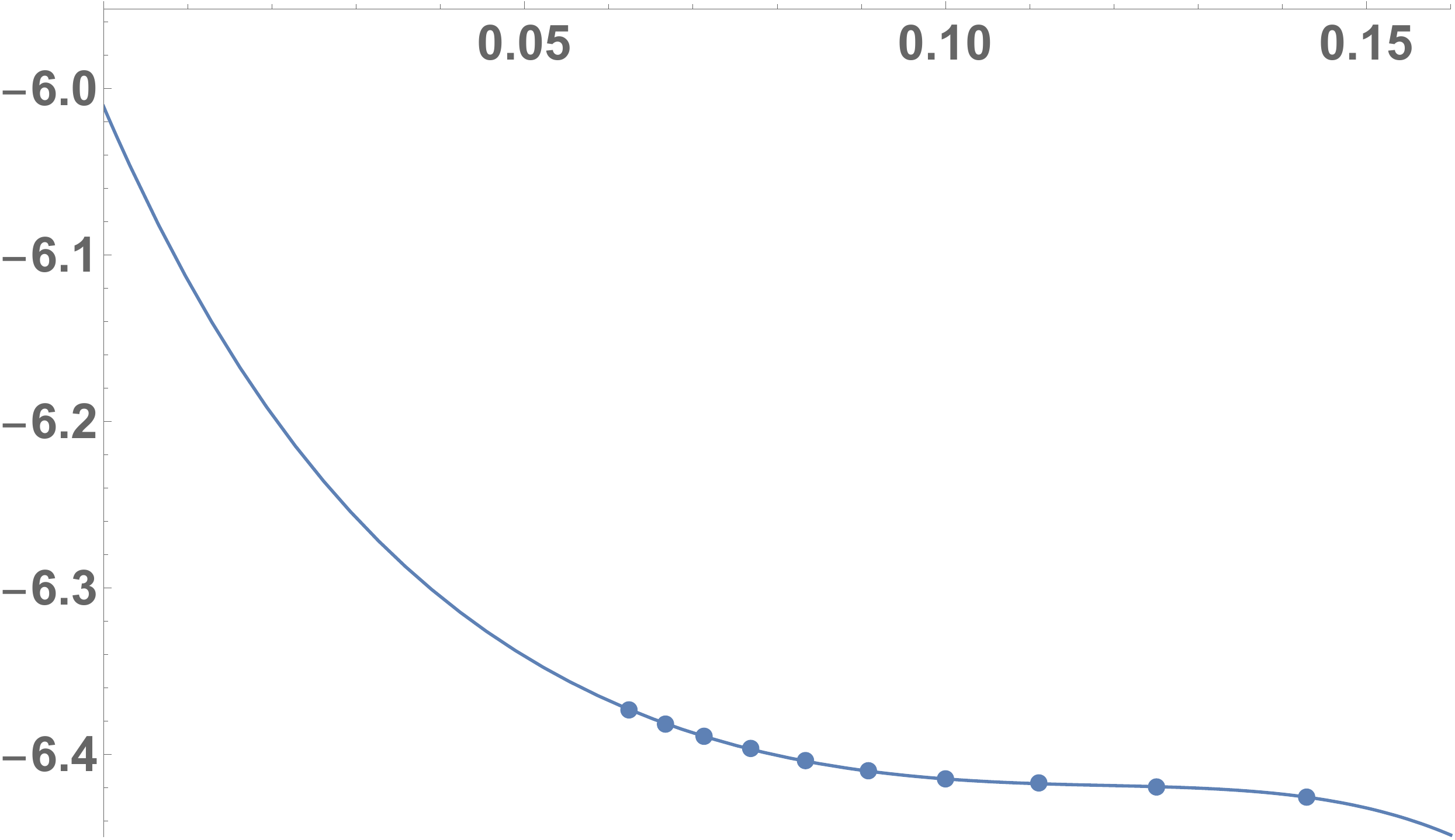} \quad
 \includegraphics[width=4.0cm]{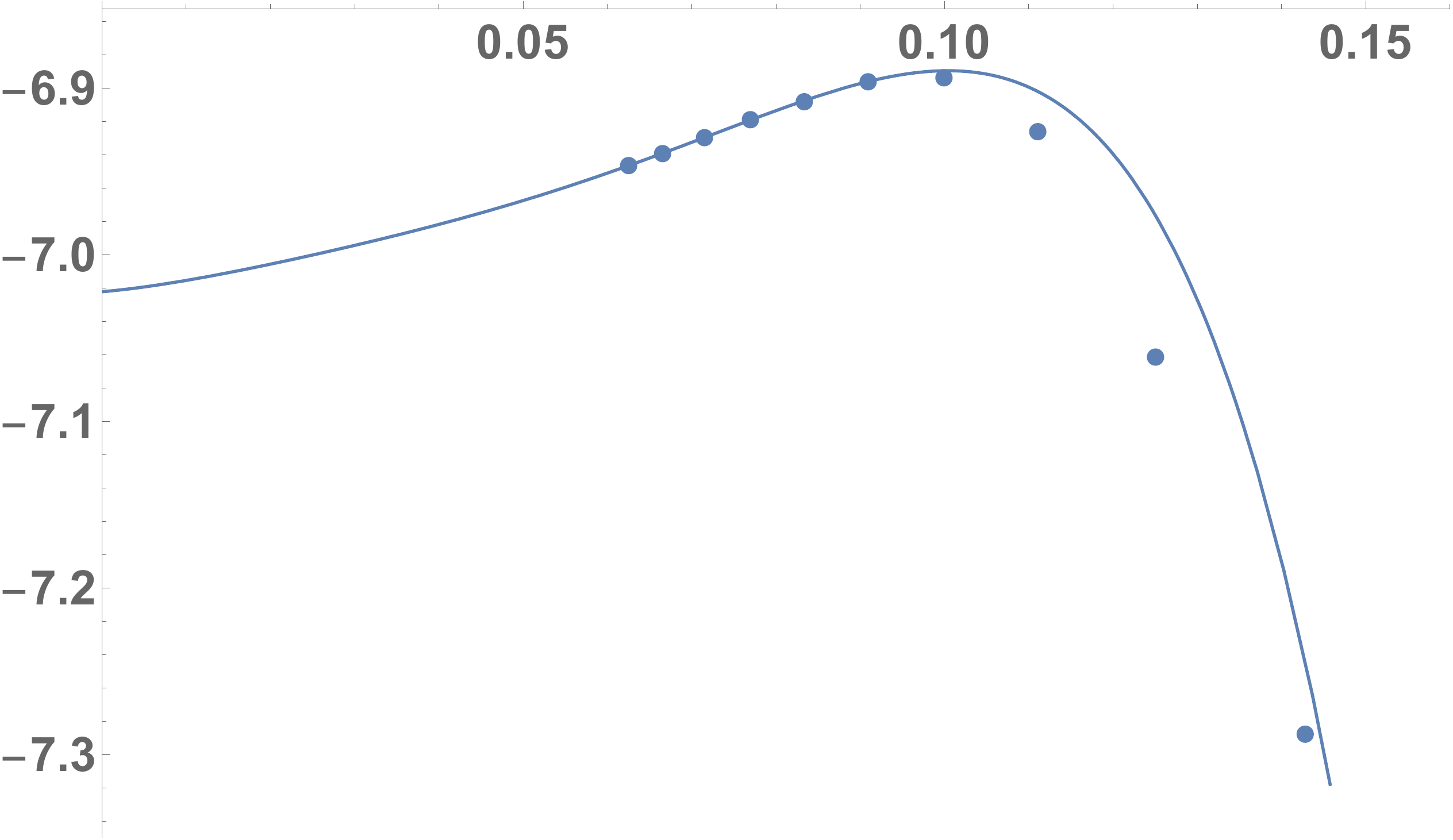}
 \tiny{($a)$} \hspace{4.0cm} \tiny{$(b)$}
 \caption{Effective scaling exponents, $\Delta_1(n)$ and $\Delta_2(n)$, for the kagome lattice, shown against $n^{-1}$ with their extrapolations.}
 \label{fig:scaling}
\end{center}
\end{figure}

To begin, we determine $\Delta_1$ from the truncated form $p_{\rm c}(n) = p_{\rm c} + A_1 n^{-\Delta_1}$. To eliminate the unknowns $p_{\rm c}$ and $A_1$,
we form the combination $q(n) \equiv \frac{p(n)-p(n-1)}{p(n-1)-p(n-2)}$. Assuming the truncated form, we have
\begin{equation*}
 q(n) = \left(1-\frac{2}{n} \right)^{\Delta_1} \frac{n^{\Delta_1} - (n-1)^{\Delta_1}}{(n-1)^{\Delta_1}-(n-2)^{\Delta_1}} \,,
\end{equation*}
a non-linear relation that determines
$\Delta_1(n)$ from three successive data points, $p_{\rm c}(n)$, $p_{\rm c}(n-1)$ and $p_{\rm c}(n-2)$. Upon supposing a reasonable starting
value, this determination is unique. Fig.\ \ref{fig:scaling}a plots $\Delta_1(n)$ against $n^{-1}$ for the kagome lattice along with a low-order polynomial fit.
Trying various orders of the fit and eliminating the lowest values of $n$ (the figure shows a fifth-order fit in $n^{-1}$ to the last seven $\Delta_1(n)$), we
conclude that $\lim_{n\to\infty} \Delta_1(n) = 6.00(2)$, cf.\ Table \ref{tab:extrap}. It appears safe to conjecture that this asymptotic value is 
$\Delta_1 = 6$ exactly for this lattice.

Next, we set $p_{\rm c}(n) = \widetilde{p}_{\rm c}(n) + \widetilde{A}_1 n^{-6}$, defining a new series $\widetilde{p}_{\rm c}(n)$, obtained from two successive
sizes, in which the leading $n^{-6}$ scaling term has been eliminated. We then repeat the above working, using now the truncated form
$\widetilde{p}_{\rm c}(n) = p_{\rm c} + A_2 n^{-\Delta_2}$. The determinations of $\Delta_2(n)$ are shown against $n^{-1}$ in Fig.\ \ref{fig:scaling}b,
and polynomial fits now lead to $\Delta_2 = 7.00(5)$, from which we conjecture that $\Delta_2 = 7$ exactly.

Table \ref{tab:extrap} compiles the scaling exponents computed in this way for the various lattices. For the first four lattices, we have data for all
$n \le 16$. For the last four, the construction of their bases $B$ requires $n$ to be even \cite{Jacobsen14}, so we have only eight data points. Obviously this
leads to more reliable determinations in the former cases.
Taken jointly, the data of Table \ref{tab:extrap} suggest that there exists two classes of non-solvable Archimedean lattices;
those (kagome, three-twelve, and snub hexagonal) for which the first three exponents are $6,7,8$,
and the remaining five lattices for which they are $4,6,8$.
We find it compelling to conjecture that the complete set of exponents are all integers $\ge 6$ for the former class, and all even integers $\ge 4$ for the latter one.

Assuming this conjecture, we can extrapolate the $p_{\rm c}(n)$ by means of the scaling form (\ref{scaling-form}). This is done by identifying a stable
compromise between the number of terms used in (\ref{scaling-form}) and the number of low-$n$ data points not included in the fits; see
Refs.\ \cite{Jacobsen15,JacobsenScullardGuttmann} for details. The end result is the central values and error bars given in Table \ref{tab:pc}.
We have checked that these are not sensitive to reasonable modifications of the values of $\Delta_k$ with $k \ge 4$, and hence do not depend
on the complete validity of the conjecture just made.

\begin{table}
\begin{center}
 \begin{tabular}{lr|l|l|l}
 Lattice & & $\Delta_1$ & $\Delta_2$ & $\Delta_3$ \\ \hline
 Kagome         & $(d)$    & 6.00(2) & 7.00(5) & 8.1(2) \\
 Four-eight      & $(e)$   & 4.00(2) & 6.0(1) & \\
 Frieze             & $(f)$ & 4.00(2) & 6.00(5) & 8.1(2) \\
 Three-twelve     & $(g)$  & 6.04(2) & 7.00(2) & 8.0(1) \\
 Cross              & $(h)$ & 4.0(2) & 6.0(5) & \\ 
 Snub square      & $(i)$  & 4.000(2) & 6.2(4) & \\
 Snub hexagonal  & $(j)$   & 5.9(1) & &  \\
 Ruby               & $(k)$ & 4.5(5) & &  \\
 \end{tabular}
 \caption{Scaling exponents for the Archimedean lattices. Blank entries cannot be determined with sufficient precision.}
 \label{tab:extrap}
\end{center}
\end{table}

\paragraph{Discussion.}
Using a parallel implementation of the eigenvalue approach to critical polynomial roots, $\simeq 3 \cdot 10^6$ CPU hours of supercomputer resources,
and a comprehensive extrapolation method, we have obtained extremely accurate values of the bond percolation thresholds (see Table~\ref{tab:pc}) on the
Archimedean lattices. In retrospect, the more naive extrapolation method used in Ref.\ \cite{Jacobsen14} now makes some of the error bars cited there
stand out as too optimistic. In the same vein, the leading scaling exponents reported in Ref.\ \cite{ScullardJacobsen16} can now be relegated to
effective exponents for small $n$. However, the $p_{\rm c}$ given in Ref.\ \cite{Jacobsen15}, based on the same scaling exponents as here, and also those of
Ref.\ \cite{ScullardJacobsen16} are fully confirmed by the more precise values now obtained. This agreement---as well as other checks, including leaving out
the $n=16$ data point from the present analysis---lends credence to the error bars given in Table~\ref{tab:pc}.

Our analysis for the scaling exponents reveals two different classes of non-solvable Archimedean lattices. This distinction cannot be explained solely from
the difference between three-fold and four-fold rotational symmetries of the lattices \cite{ScullardJacobsen16}, although it certainly is remarkable that all members of the first class
(kagome, three-twelve, snub hexagonal) enjoy three-fold symmetries. The proposed scaling exponents are compatible with conformal field theory
predictions \cite{Jacobsen15}, but it remains unclear how to derive them analytically.

In addition to the interesting open problems related to the critical polynomial method, it is yet to be determined how widely applicable it is. The definition (\ref{eq:P_B}) is easily generalized to the $Q$-state Potts model and gives excellent estimates for any $Q$ \cite{JacobsenScullard2013,ScullardJacobsen16}, even in the imaginary temperature regime \cite{Saleur91}. The eigenvalue method presented here was also adapted to compute the growth constant of self-avoiding walks and was able finally to rule out a longstanding conjecture \cite{JacobsenScullardGuttmann}. Generalizations to site percolation, coupled Potts models, or to non-planar models, are currently under investigation.

\paragraph{Acknowledgements.}
We thank Robert Ziff for sharing his numerical bond threshold of the snub hexagonal lattice reported in Table \ref{tab:pc}. The work of CRS was performed under the auspices of the U.S. Department of Energy at the Lawrence Livermore National Laboratory under Contract No.\ DE-AC52-07NA27344.
JLJ is grateful for support from the European Research Council under the Advanced Grant NuQFT.
This work was granted access to the HPC resources of IDRIS under the allocation 2016-057751 attributed by GENCI (Grand Equipement National de Calcul Intensif).

\bibliography{JS.bib}
\end{document}